\documentclass[12pt,preprint]{elsarticle}
\usepackage[active]{srcltx} 
\usepackage{hyperref}
\usepackage{graphics}
\usepackage{epsfig}
\usepackage{amsmath}
\usepackage{mathtools}
\usepackage[english]{babel}
\usepackage{graphicx}
\usepackage{url}
 \usepackage{footnote}
\newcommand{\eq}[1]{\begin{align} #1 \end{align}}
\RequirePackage{lineno}
\makeatletter
\def\ps@pprintTitle{%
 \let\@oddhead\@empty
 \let\@evenhead\@empty
 \def\@oddfoot{}%
 \let\@evenfoot\@oddfoot}
\makeatother

\begin{document}

\begin{frontmatter}
\title{A model-free procedure to correct for volume fluctuations in E-by-E analyses of particle multiplicities}  

\author[GSI]{Anar Rustamov}
\author[IKF,GSI,HFHF]{Joachim Stroth}
\author[GSI]{Romain Holzmann}
\address[GSI]{GSI Helmholtzzentrum f\"ur Schwerionenforschung GmbH, 64291 Darmstadt, Germany}
\address[IKF]{Institut f\"ur Kernphysik, Goethe-Universit\"at, 60438 Frankfurt am Main, Germany}
\address[HFHF]{Helmholtz Forschungsakademie
Hessen für FAIR, Campus Frankfurt,\\ 60438 Frankfurt am Main, Germany}
\begin{abstract}
We develop an innovative and unbiased procedure, based on event mixing, to account for unavoidable contributions from volume (or system size) fluctuations to experimentally measured moments of particle multiplicity distributions produced in relativistic nuclear collisions. Within the wounded-nucleon model they are characterized by fluctuations of the number of wounded nucleons, the latter usually referred to as participants. For the first time we extract participant fluctuations directly from the data used for the fluctuation analysis, i.e., without involving model calculations. To achieve this we constructed a dedicated event-mixing algorithm that eliminates all possible correlations between produced particles while preserving the volume fluctuations. The procedure provides direct access to the cumulants of wounded-nucleon distributions, which can be used to account for non-critical contributions to the experimentally measured cumulants of multiplicity distributions. 
\end{abstract}
     
\date{\today}  

\end{frontmatter}

\section{Introduction}
Fluctuations of conserved charges such as baryon number ($B$), electric charge ($Q$), strangeness ($S$) is the focus of current experimental and theoretical investigations of relativistic nuclear collisions worldwide~\cite{Stephanov:1998dy,Bazavov:2020bjn, Rustamov:2022hdi, Braun-Munzinger:2022bkc}. 
They are identified as promising probes of the Equation of State (EoS) of a system under the study through its response to infinitesimal changes in external parameters, thereby providing direct access to the phase structure of the system.  
At center-of-mass energies available at LHC down to about 12\;GeV per nucleon pair recent theoretical calculations imply that chiral symmetry is restored in a smooth crossover transition~\cite{HotQCD:2018pds, Almasi:2017bhq, Borsanyi:2018grb}. 
In the region of high net-baryon density, where systematic lattice QCD calculations are not directly applicable,  effective model calculations suggest that the strongly interacting matter undergoes a first order chiral phase transition~\cite{Asakawa:1989bq, Rajagopal:1992qz, Stephanov:1998dy, Stephanov:2006dn, Sasaki:2007db}. 
The conjectured chiral critical end point (CEP), at which the matter exhibits a second order phase transition, should terminate the anticipated first order chiral phase transition line. 
Extensive experimental efforts to measure net-proton number fluctuations are underway~\cite{Rustamov:2017lio, ALICE:2022xpf, ALICE:2019nbs, HADES:2020wpc, STAR:2020tga, Adhikary:2022sdh, Rustamov:2020ekv}, in order to verify these theoretical predictions of a smooth crossover and to locate the conjectured CEP in the phase diagram of strongly interacting matter. 

Within the Grand Canonical Ensemble (GCE) of statistical mechanics net-charge fluctuations can be quantified with generalized susceptibilities. 
In a thermal system of fixed volume $V$ at a temperature $T$ the $n^{th}$ order susceptibilities $\chi_{q}^{n}$ are defined as derivatives of the logarithm of partition function $Z$ with respect to the chemical potentials $\mu_{q}$ responsible for the conservation of the corresponding charge $q \in \{B, Q, S\}$ on average, and evaluated at vanishing values of chemical potentials~\cite{Rustamov:2022hdi, Braun-Munzinger:2022bkc}:
\begin{equation}
\label{eq-chi}
    \chi_{n}^q \equiv \frac{1}{VT^{3}}\left.\frac{\partial^{n}ln Z(T, V, \mu_{B}, \mu_{Q}, \mu_{S})}{\partial \hat{\mu}_{q}^n}\right\rvert_{\vec{\mu} = 0}  = \frac{1}{VT^{3}}\kappa_{n}(N_{q})\;.
\end{equation}
Here $\hat{\mu}_{q}=\mu_{q}/T$ and $\kappa_{n}(N_{q})$ is the $n^{th}$ order cumulant of net-charge number distribution $N_{q}$, which can be measured in experiments. 
Other parameters of the system, such as pressure, entropy density etc., can be calculated by taking appropriate derivatives of $ln(Z)$.

In principle Eq.~\ref{eq-chi} already establishes a direct link between experimentally measured cumulants $\kappa_{n}(N_{q})$ and theoretically evaluated susceptibilities $\chi_{n}^q$, thus allowing to probe the EoS of the system, which is encoded in the partition function $Z$. Clearly the experimental measurements are to be done in a subspace of the full phase space, yet keeping in mind that in the full phase space there are no fluctuations of conserved charges, i.e., they are conserved in each event. 
And this is exactly what is missing in Eq.~\ref{eq-chi}. 
The solution to this problem is provided by exploiting the Canonical Ensemble of statistical mechanics in the full phase space to account for event-by-event net-charge conservation effects inside the selected subspace~\cite{Nahrgang:2009dqc, Bzdak:2012an,  Braun-Munzinger:2020jbk, Braun-Munzinger:2018yru, Braun-Munzinger:2019yxj}. 

However, for a proper comparison of experimental results with theory predictions several additional contributions to the experimental measurements need to be controlled with high precision~\cite{Braun-Munzinger:2016yjz}. 
Among those are fluctuations due to the statistical nature of the detector response~\cite{Bzdak:2013pha,Bzdak:2016qdc,Arslandok:2018pcu,Luo:2017faz, Rustamov:2012bx, Gazdzicki:2019rrq}, effects stemming from detecting only proxies (e.g.\ protons) of the true baryon number fluctuations~\cite{Kitazawa:2012at,Luo:2017faz}, a mitigation of the critical signal due to residual dynamics~\cite{Asakawa:2015ybt, Vovchenko:2021kxx,Vovchenko:2022szk}, and, last but not least, an insufficient control of the volume associated with the particle-emitting source.
In this paper, we will address the rather complicated problem of volume fluctuations for which there is so far no solution proven to work also at low collision energies.
Equation~\ref{eq-chi} is valid only if the system volume is fixed, a condition which is hard, if not impossible, to achieve in experiments where the volume is not directly observable. 
Commonly, in particular at high collision energies where the colliding nuclei are sufficiently Lorentz contracted in the center-of-mass, the volume is associated with the number of wounded nucleons~\cite{Bialas:1976ed}, usually referred to as participants~\cite{Braun-Munzinger:2016yjz}. 
In fact, experimental data are analyzed in centrality percentiles corresponding to $n\%$ most central collisions by introducing selection criteria e.g.\ in the energy deposited in a forward calorimeter or the multiplicity of charged particles, either in the full acceptance or in its sub-ranges etc.~\cite{ALICE:2013hur, HADES:2017def}. 
For the latter, care must be taken to ensure that the evaluated particles are not simultaneously used to determine the critical fluctuations~\cite{STAR:2020tga}.
By fitting the so obtained distributions with the Glauber Monte Carlo Model~\cite{Loizides:2014vua}, where in addition it is assumed that particles are produced from statistically independent sources, the distributions of participants corresponding to a given centrality class are obtained.
Consequently, for each centrality class the number of participants fluctuates from event to event, which prevents direct exploitation of Eq.~\ref{eq-chi}~\cite{Braun-Munzinger:2016yjz}. 

To date, three conceptually distinct approaches have been developed to circumvent this crucial experimental artifact in measurements of event-by-event particle number fluctuations: (i) By evaluating contributions from volume~\cite{Skokov:2012ds} or participant~\cite{Braun-Munzinger:2016yjz} fluctuations within the model of independent sources for particle production. 
The latter approach, however, needs probability distributions of participants as input. 
In~\cite{Braun-Munzinger:2016yjz} a detailed procedure is worked out to account for participant fluctuations using the Glauber Monte Carlo model. 
(ii) Introducing the so-called strongly intensive quantities which, by a specific construction, eliminate participant fluctuations~\cite{Gorenstein:2011vq}. This method also assumes statistically independent particle production.  
(iii) Using an unfolding algorithm which also relies on model calculations~\cite{Esumi:2020xdo}. 

In this work we propose, as a radically new approach, to reconstruct the participant fluctuations directly from the data used for the fluctuation analysis. 
This is achieved by constructing a dedicated algorithm for event mixing which removes all correlations between particles while keeping the participant fluctuations unaffected. 
The paper is organized in the following way. 
In section~\ref{l_def} we present definitions and notations. 
The main idea behind the method is introduced in sections~\ref{l-fact} and~\ref{l-mixing}, in which the necessary analytic expressions are given and the procedure for event mixing is introduced. 
The results obtained are presented in section~\ref{l_res} and conclusions are given in section~\ref{l_conc}.
\section {Definitions and notations used in the paper}
\label{l_def}
In the following we use the notations for cumulants introduced in~\cite{Braun-Munzinger:2016yjz}.
The $r^{th}$ central moment of a discrete
random variable $X$, with probability distribution $P(X)$, is
defined as
\begin{equation}
\mu_{r} \equiv  \left<\left(X-\left<X\right>\right)^{r}\right> = \sum_{X}\left(X-\left<X\right>\right)^{r}P(X),\\
\label{cum_definition1}
\end{equation}
where $\left<X\right>$ denotes the mean of the distribution
\begin{equation}
\left<X\right> =  \sum_{X}XP(X) .\\
\label{mean_definition1}
\end{equation}
In a similar way, we introduce moments about the origin, thereafter referred to as raw moments 
\begin{equation}
 \left<X^{r}\right> = \sum_{X}X^{r}P(X).\\
\label{mean_definition2}
\end{equation}
The cumulants of $X$ are defined as the coefficients in the Maclaurin
series of the logarithm of the characteristic function of $X$.  
The first four cumulants read
 \eq{\label{kumulants_definition}
  &\kappa_{1} = \left<X\right> , \nonumber\\ &\kappa_{2} = \mu_{2} =
  \left<X^{2}\right> - \left<X\right>^{2} , \nonumber\\ &\kappa_{3} =
  \mu_{3} = \left<X^{3}\right> - 3\left<X^{2}\right>\left<X\right> +
  2\left<X\right>^{3}, \\ &\kappa_{4} = \mu_{4} - 3\mu_{2}^{2} =
  \left<X^{4}\right> - 4\left<X^{3}\right>\left<X\right> -
  3\left<X^{2}\right>^{2} \nonumber\\ &
  +12\left<X^{2}\right>\left<X\right>^{2} -
  6\left<X\right>^{4}. \nonumber }
For a random variable $X$, following the Poisson probability distribution $P(X=k;\lambda) = e^{-\lambda}\lambda^{k}/k!$, any order cumulants of its distribution are equal to its mean, that is $\kappa_{n}(X)=\lambda$.
\section{Factorization approach}
\label{l-fact}
The cumulants $\kappa_{n}$ reconstructed from the multiplicity distributions contain both a signal component and a background contribution due to the fluctuations of the participants. 
The ultimate goal of the analysis is to disentangle the signal part from non-critical contributions before they are compared to the theoretical calculations using e.g.\ Eq.~\ref{eq-chi}. 
Contributions from fluctuations of participants are usually accounted for by exploiting statistically independent sources for particle production~\cite{Braun-Munzinger:2016yjz}, such as the Wounded Nucleon Model (WNM)~\cite{Bialas:1976ed}.
Within WNM, the analytical expressions for the cumulants of (net-)particle distributions can be obtained. 
The first four cumulants read~\cite{Skokov:2012ds, Braun-Munzinger:2016yjz}:
\eq{
\label{cum1}
&\kappa_{1}(N) = \left<N_{W}\right>\kappa_{1}(n) = \left<N_{W}\right>\left<n\right>,\\
\label{cum2}
&\kappa_{2}(N) = \left<N_{W}\right>\kappa_{2}(n) + \left<n\right>^{2}\kappa_{2}(N_{W}),\\
\label{cum3}
&\kappa_{3}(N) =\left<N_{W}\right> \kappa_{3}(n) +   3\left<n\right>\kappa_{2}( n)\kappa_{2}(N_{W}) + \left<n\right>^{3}\kappa_{3}(N_{W}),\\
\label{cum4}
&\kappa_{4}(N) = \left<N_{W}\right>\kappa_{4}(n) +
4\left<n\right>\kappa_{3}(n)\kappa_{2}(N_{W}) \\ &+
3\kappa_{2}^{2}(n)\kappa_{2}(N_{W}) + 6\left<n\right>^{2}\kappa_{2}(n)\kappa_{3}(N_{W}) + \left<
n\right>^{4}\kappa_{4}(N_{W}),\nonumber }
where $\kappa_{n}(n)$, $\kappa_{n}(N)$ denote the cumulants of particles from a single wounded nucleon (or source) and those obtained after averaging over the wounded-nucleon distribution, respectively. 
Within the same framework of statistically independent particle sources we obtain the following expression for the covariance between the multiplicities of two distinct particle species $N_{1}$, $N_{2}$:
\begin{equation}
cov(N_{1},N_{2}) = \left<N_{W}\right>cov(n_{1}, n_{2})+\left<n_{1}\right>\left<n_{2}\right>\kappa_{2}(N_{W}),
\label{eq_cov}
\end{equation}
with $cov(n_{1}, n_{2})$ and $cov(N_{1}, N_{2})$ denoting the covariances due to emission from a single source and after averaging over the wounded-nucleon distribution, respectively. 
Obviously, \emph{a priory} the values  of $k_{n}(n)$ and $cov(n_{1}, n_{2})$ are unknown -- they actually represent the physics we are after. 
In order to isolate them, as it is evident from Eqs.~\ref{cum1}-~\ref{eq_cov}, we need additional information, namely distributions of wounded nucleons or rather the cumulants of wounded-nucleon distributions $k_{n}(N_{W})$. 
That is where the model dependence sets in. 
Indeed, nearly in all experiments the cumulants of wounded nucleons are obtained by exploiting models, typically Glauber Monte Carlo simulations or transport codes. 
The latter are known to introduce biases, in particular at low collision energies~\cite{STAR:2021fge}.
On the other hand, at beam energies below 3\,$A$\,GeV the build-up of transverse energy and momentum in the early stage of the collision, when projectile- and target-like spectator volumes are not yet separated from the participant zone, can lead to participant formation outside the geometrical overlap volume.    
In that case, a strict scaling of the number of participants, as the sources of particle emission, with the size of the fireball volume cannot be assumed anymore.
\section{Model independent extraction of participant fluctuations}
\label{l-mixing}
\begin{figure}[!htb]
    \centering
    \includegraphics[width=1.\linewidth,clip=true]{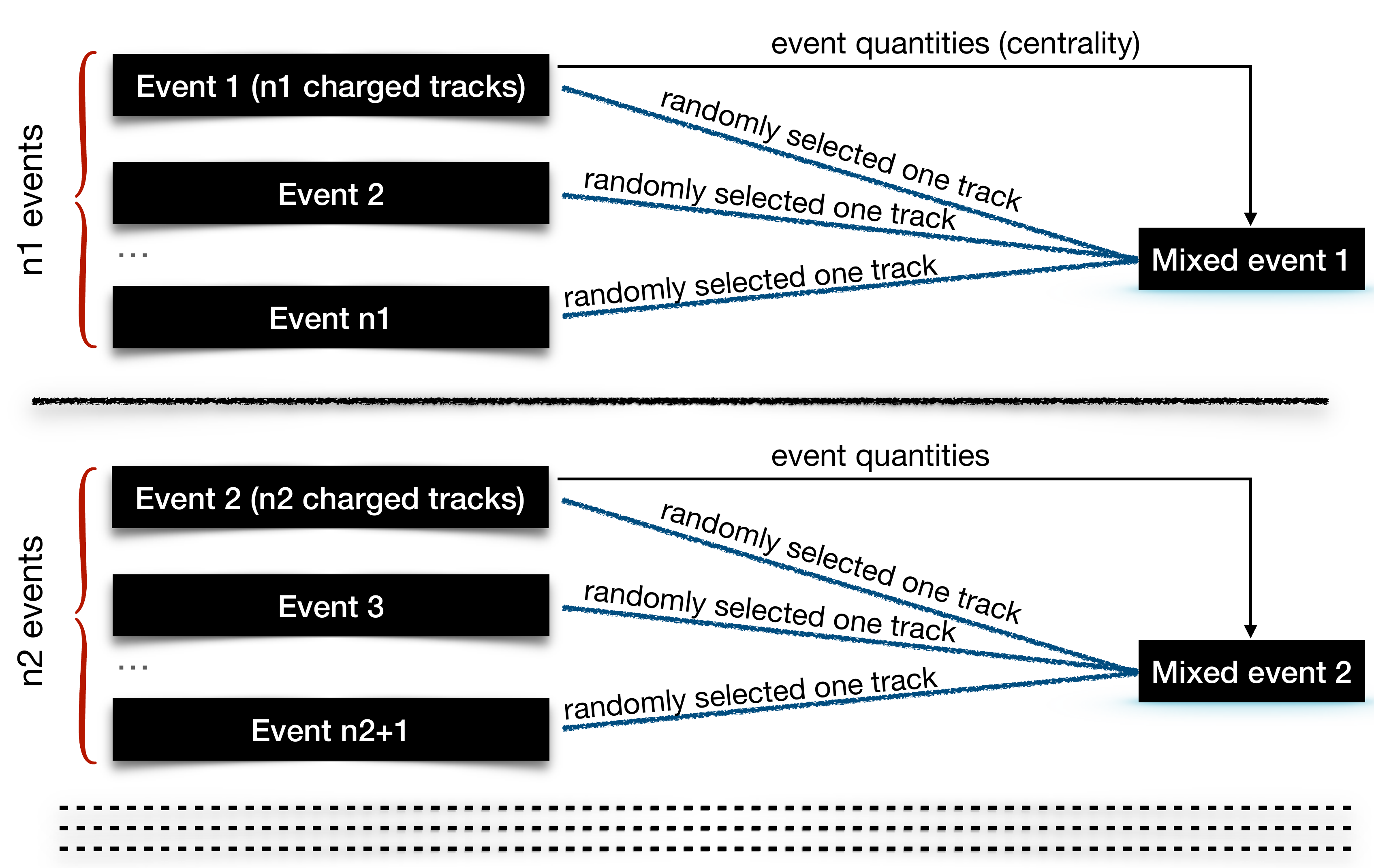}
    \caption{The strategy for event mixing used to remove correlations between particles while preserving participant fluctuations.}
    \label{fig:emixing}
\end{figure}
The method we develop acts differently. In a first step we mix the events in a way to preserve the fluctuations of wounded nucleons but remove all correlations between particles; the procedure is illustrated in Fig.~\ref{fig:emixing}. For a given experimental event we count the number of all reconstructed charged particles $n_{ch}$. Next, we select randomly one charged track from that event, while the rest of the charged particles are each randomly taken from the following $n_{ch} - 1$ events, one per each event. In this way, the newly created mixed event contains $n_{ch}$ particles, but the composition of particle species in the mixed event is randomized. The procedure is repeated event-by-event until a desired statistics of mixed events is reached. The correlations between all particle species in mixed events are completely eliminated,  yet the original wounded-nucleon fluctuations are preserved. This is because the total number of charged particles in real and mixed events is kept to be the same. This also implies that in Eq.~\ref{eq_cov} the single-source covariance term vanishes, yielding: 

\begin{equation}
cov(N_{1},N_{2}) = \left<n_{1}\right>\left<n_{2}\right>\kappa_{2}(N_{W}) = \left<N_{1}\right>\left<N_{2}\right>\frac{\kappa_{2}(N_{W})}{\left<N_{W}\right>^{2}},
\label{mix_1}
\end{equation}

In a similar way, using Eq.~\ref{cum2} and realizing that for mixed events $\kappa_{n}(n) = \left<n\right>$ (Poisson-distributed particles, because fully uncorrelated), we obtain the following expression:

\begin{equation}
\kappa_{2}(N) = \left<N_{W}\right>\left<n\right>+\left<n\right>^{2}\kappa_{2}(N_{W}) = \left<N\right>+\left<N\right>^{2}\frac{\kappa_{2}(N_{W})}{\left<N_{W}\right>^{2}},
\label{mix_2}
\end{equation}

Using Eg.~\ref{mix_1}, the second cumulants of participant fluctuations are obtained from covariances between multiplicities of different particle species $N_{1}$ and $N_{2}$:

\begin{equation}
\frac{\kappa_{2}(N_{W})}{\left<N_{W}\right>^{2}} = \frac{cov(N_{1},N_{2})}{\left<N_{1}\right>\left<N_{2}\right>} 
\label{mix_3}
\end{equation}

In a similar way, using Eq.~\ref{mix_2}, participant fluctuations can be extracted from the reconstructed second order cumulants for particle type $N$: 

\begin{equation}
\frac{\kappa_{2}(N_{W})}{\left<N_{W}\right>^{2}} = \frac{\kappa_{2}(N)}{\left<N\right>^{2}} - \frac{1}{\left<N\right>}
\label{mix_4}
\end{equation}

The covariances and cumulants entering the right hand sides of Eqs.~\ref{mix_3} and~\ref{mix_4} are to be evaluated using mixed events.  Note that,
by exploiting Eqs.~\ref{cum3} and~\ref{cum4}, higher-order cumulants of participant distributions can also be obtained.

\section{Results and discussions}
\label{l_res}

\begin{figure}[!htb]
    \centering
    \includegraphics[width=.49\linewidth,clip=true]{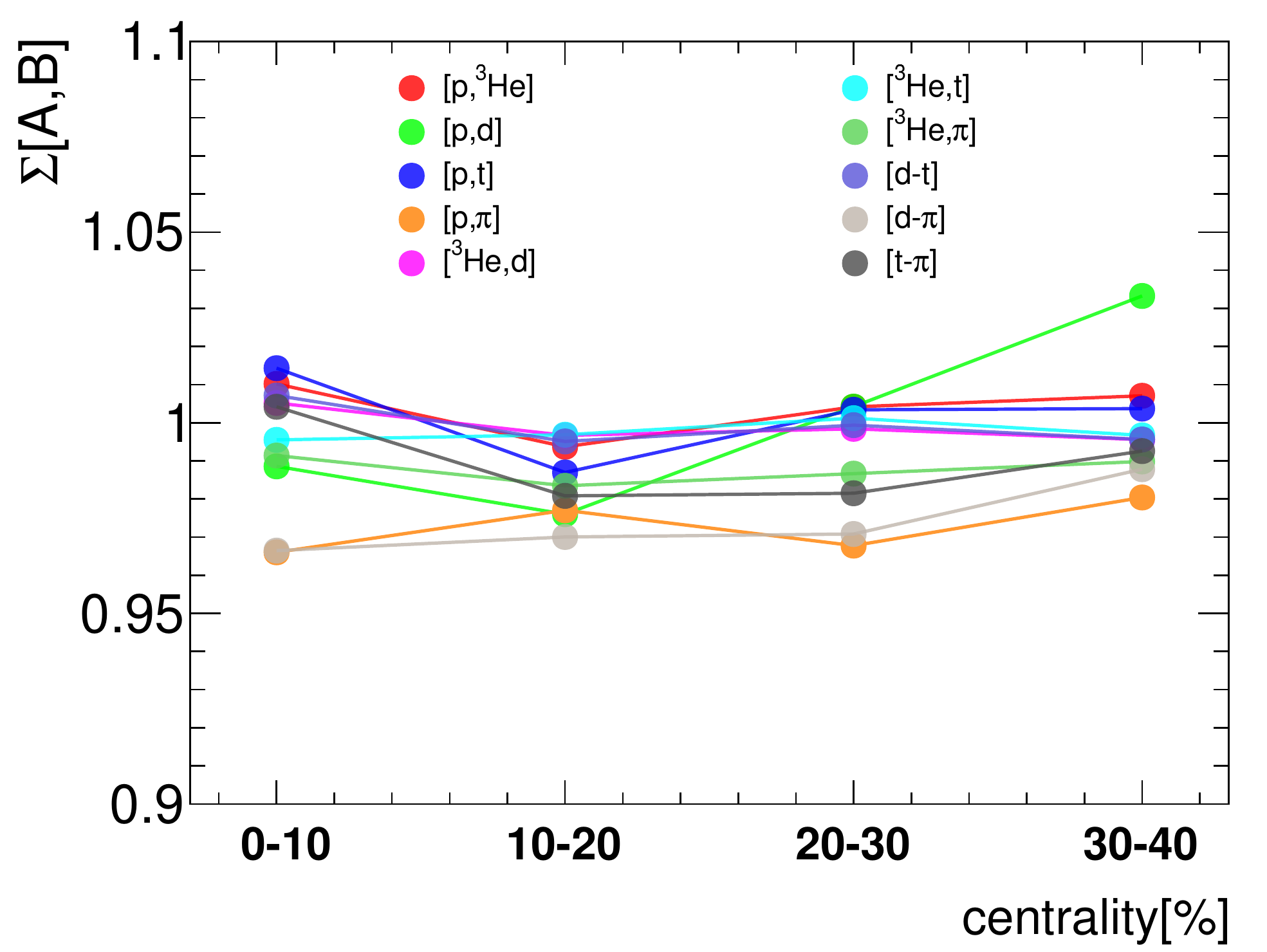}
    \includegraphics[width=.49\linewidth,clip=true]{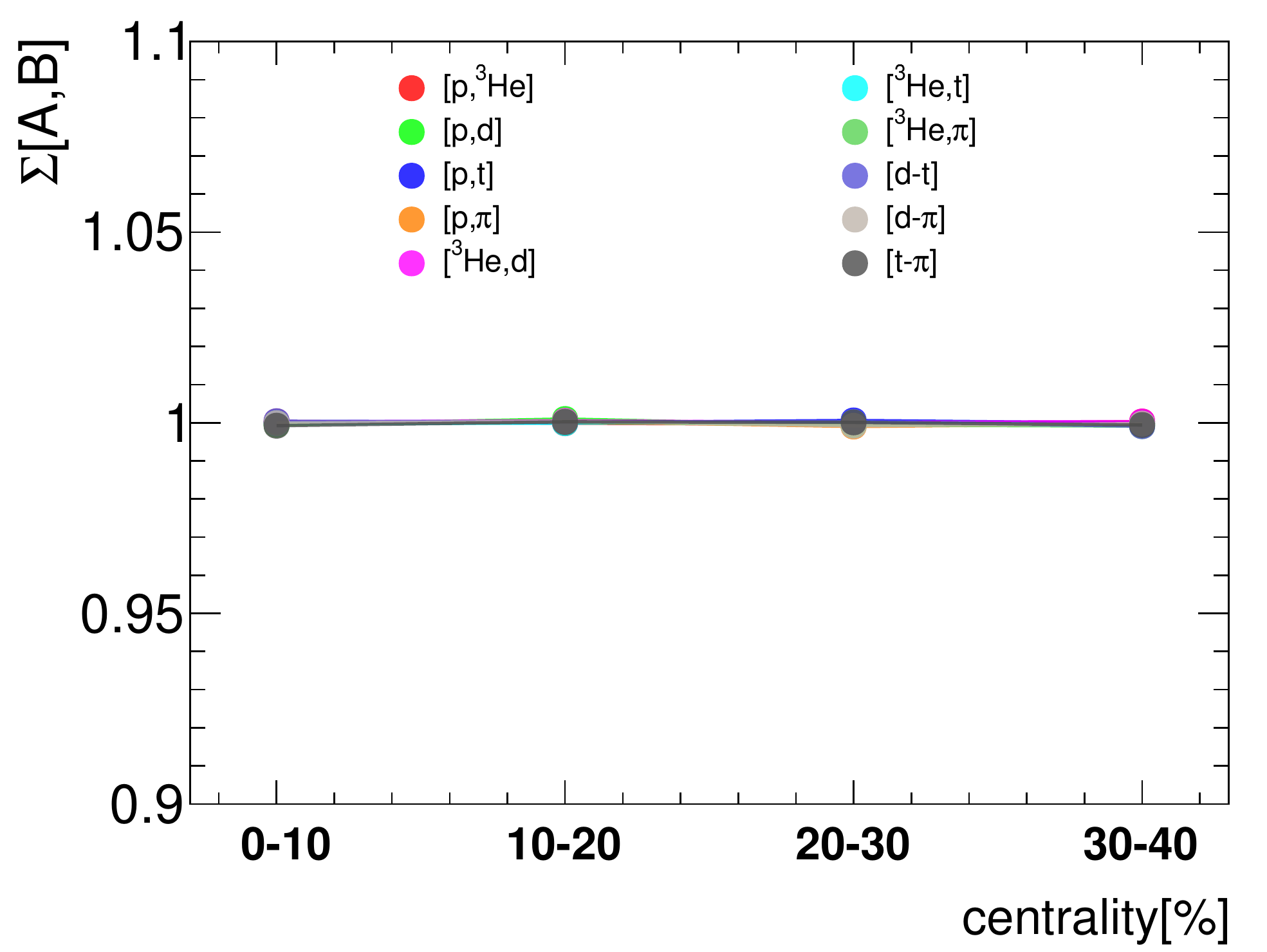}    
    \caption{Fluctuation measure $\Sigma[A, B]$ reconstructed for 10 different particle pairs, indicated with colored symbols, as a function of collision centrality for original (left panel) and mixed (right panel) IQMD events.  }
    \label{fig:SIGMA}
\end{figure}

The developed procedure is tested on events generated for Au-Au collisions at $\sqrt{s_{\mathrm{NN}}}$ = 2.4 GeV with the IQMD transport model~\cite{Hartnack:1997ez, Gossiaux:1997hi}. Cumulants  of multiplicity distributions are reconstructed for five different particle species: proton (p), pion ($\pi$), deuteron (d), triton (t) and nucleus of $He$ isotope ($^{3}He$). Moreover, covariances are reconstructed between multiplicites of all possible combinations of particle pairs: [p, $^{3}He$], [p, d], [p, t], [p, $\pi$], [$^{3}He$, d], [$^{3}He$, t], [$^{3}He$, $\pi$], [d, t], [d, $\pi$], [t, $\pi$]. The statistical uncertainties are estimated using the subsampling approach~\cite{ALICE:2019nbs, HADES:2020wpc}. The analysis is performed  in four different centrality classes of 0-10\%, 10-20\%, 20-30\%, and 30-40$\%$, selected by applying appropriate windows in the impact parameter distribution provided by IQMD.   Alternatively one could apply different centrality selection criteria to meet experimental centrality selection methods, however for the results presented in this section this is immaterial. The method presented works for any kind of centrality selection used in experiments. 

A sample of mixed events is generated as prescribed in section~\ref{l-mixing}, further taking care that only events falling into a given centrality class are used\footnote{We cross-checked, that mixing  events from different centrality classes does not introduce significant biases.}. In addition, only charged particles within kinematic region delimited as $|y|<0.4 $ and $0.4<p_{T}<$ 1.6 GeV/c are used, with $y$ and $p_{T}$ referring to the rapidity (using proton mass) and transverse momentum of charged particles. In order to verify that in the mixed events all correlations between particle multiplicities are lifted, we present in Fig.~\ref{fig:SIGMA} the centrality dependence of the strongly intensive fluctuation measure $\Sigma[A,B]$, estimated for 10 different particle pairs [A, B] mentioned above. Within WNM it depends neither on the mean number of participants (centrality, system size or system volume) nor on its fluctuations, and is defined as~\cite{Gorenstein:2011vq}:

\begin{equation}
\Sigma[A,B] = \frac{\left<A\right>\omega(B)+\left<B\right>\omega(A)-2cov(A,B)}{\left<A+B\right>},
\label{eq_SIGMA}
\end{equation}
where $\omega(A)$ and $\omega(B)$ are scaled variances of multiplicity distributions of particle species $A$ and $B$, respectively, and $cov(A,B)$ is the covariance between the corresponding multiplicity distributions of particles $A$ and $B$, $\left<A\right>$ and $\left <B\right>$ in Eq.~\ref{eq_SIGMA} stand for event-averaged mean multiplicities. It is evident from Eq.~\ref{eq_SIGMA}, that in the case of missing correlations between multiplicities of particles $A$ and $B$, the fluctuation measure $\Sigma[A,B]$ becomes unity. Indeed in this case $cov(A,B)$ vanishes and both scaled variances $\omega(A)$ and $\omega(B)$  become unity.

\begin{figure}[!htb]
    \centering
    \includegraphics[width=.49\linewidth,clip=true]{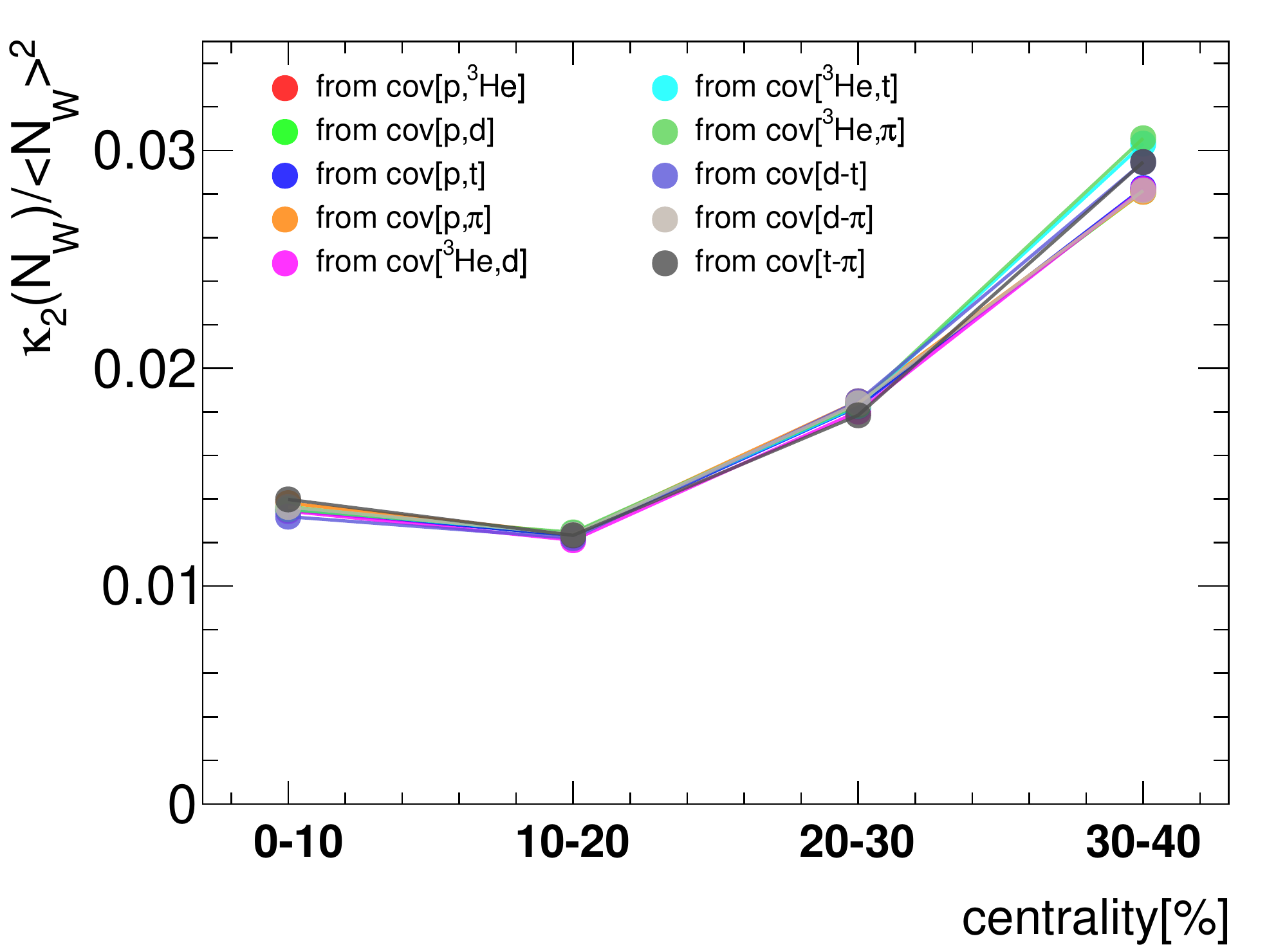}
    \includegraphics[width=.49\linewidth,clip=true]{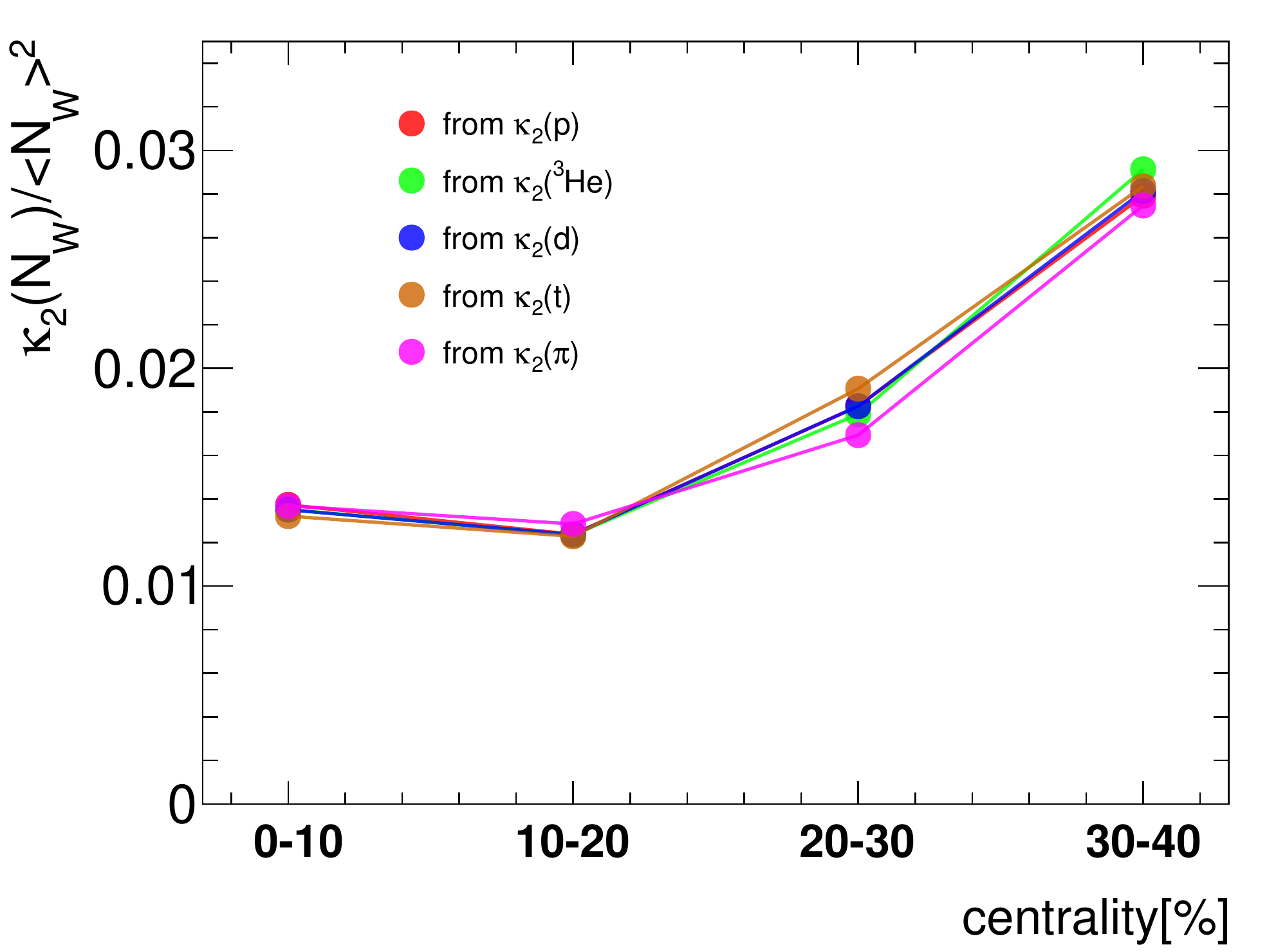}
    \caption{Normalized second order cumulants of participants as obtained from covariances (left panel) and single particle cumulants (right panel).}
    \label{fig:k2nw}
\end{figure}

\begin{figure}[!htb]
    \centering
    \includegraphics[width=1.\linewidth,clip=true]{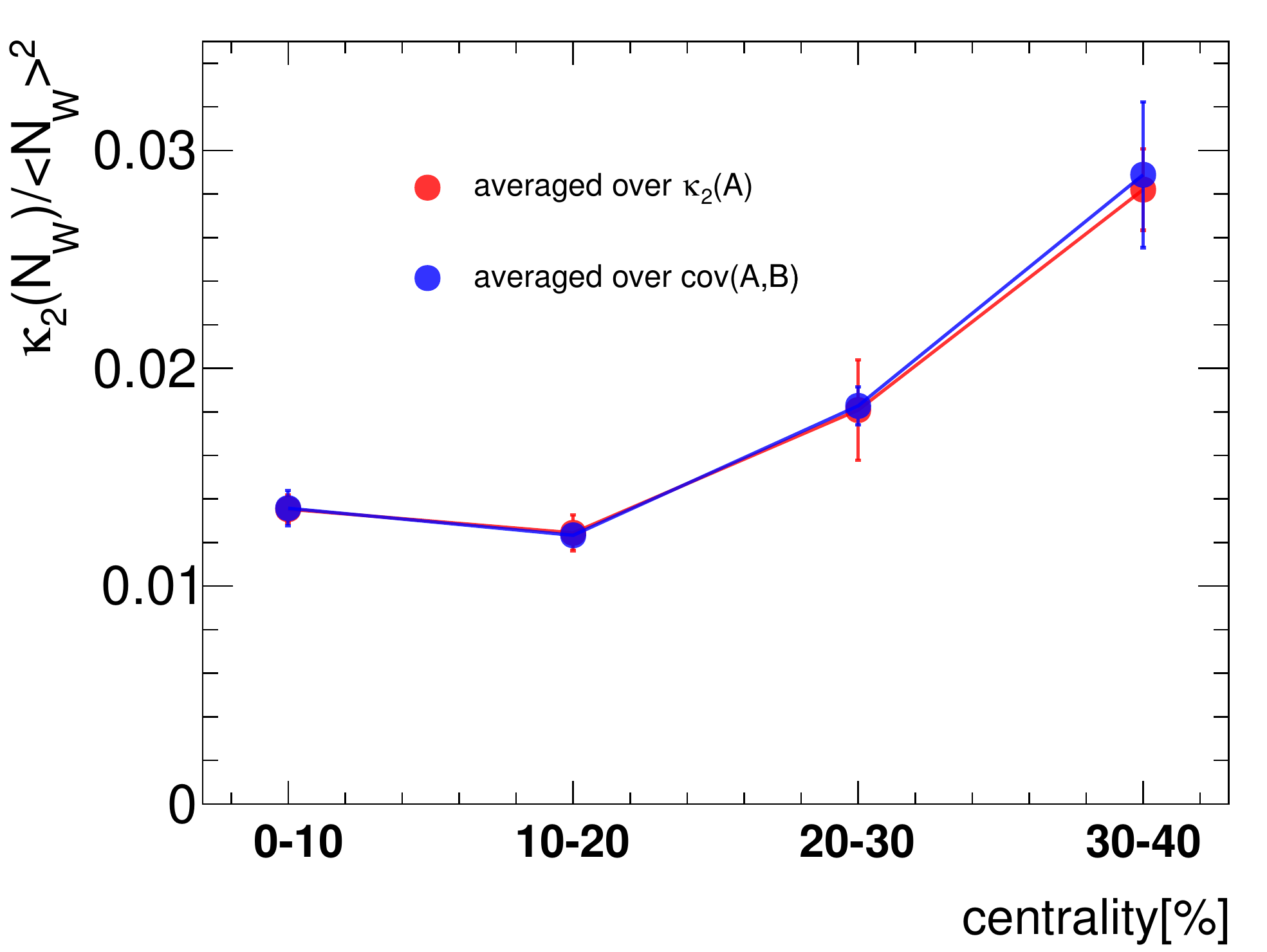}
    \caption{Normalized second order cumulants of participants  obtained from single particle cumulants averaged over five different particle species (red symbols) and those extracted from the covariances averaged over all combinations between all particle pairs (blue symbols).}
    \label{fig:k2nw_aver}
\end{figure}

The left panel of Fig.~\ref{fig:SIGMA} shows the centrality dependence of $\Sigma[A,B]$ for real (original) events while in the right panel similar results for the mixed events are presented. Inspection of  Fig.~\ref{fig:SIGMA} clearly demonstrates that in mixed events the values of $\Sigma[A,B]$ are all becoming unity, hence indicating that any possible correlations between multiplicities of every particle pair are eliminated. After this verification procedure we use  mixed events to extract the fluctuations of participants. Hence we turn back to quantities which do depend on fluctuations of participants, i.e., cumulants of multiplicity distributions and covariances between them. In the following, only  mixed events are used. We first reconstruct covariances between multiplicity distributions of all particle pairs and using Eq.~\ref{mix_3} extract the normalized second order cumulant of participant distributions $\kappa_{2}(N_{W})/\left<N_{W}\right>^{2}$. Obtained results as a function of centrality are presented in the left panel of Fig.~\ref{fig:k2nw}. Alternatively, we obtain the same information (cf. right panel of Fig.~\ref{fig:k2nw}) using Eq.~\ref{mix_4} and reconstructed  single particle cumulants, $\kappa_{2}(A)$. Fig.~\ref{fig:k2nw} demonstrates, that the $\kappa_{2}(N_{W})/\left<N_{W}\right>^{2}$ values extracted from covariances and single particle cumulants are in reasonable agreement with a mild dependence on particle species, thus demonstrating the robustness of the method. This is clearly visible in Fig.~\ref{fig:k2nw_aver}, where the centrality dependence of the $\kappa_{2}(N_{W})/\left<N_{W}\right>^{2}$ values obtained from single particle cumulants averaged over five different particle species (red symbols) and those extracted from the covariances averaged over all combinations between all particle pairs (blue symbols) are presented. 

\section{Conclusion}
\label{l_conc}
In summary, we present a model-independent approach to disentangle non-critical contributions to experimentally measured particle number fluctuations. Specifically we demonstrate a possibility of eliminating contributions from participant fluctuations based on measurements only, i.e., without involving the models. The method was tested on events generated at $\sqrt{s_{\mathrm{NN}}}$ = 2.4 GeV  with the IQMD transport model. The participant fluctuations are extracted in two different ways: (i) Using second order cumulants of five different  particles species and (ii) From covariances between multiplicity distributions of all possible pairs of particles. Both approaches yield similar results, demonstrating that the method developed is robust enough. The proposed method is essential for nuclear collisions at low collisions energies performed at GSI/SIS and BNL/RHIC, but can be applied at LHC energies as well. Furthermore, the method is general enough to allow reconstructing any higher-order cumulant of the participant distribution.

\section{Acknowledgements} 
AR acknowledges stimulating discussions with Peter Braun-Munzinger and Johanna Stachel.

\bibliography{VolumeFluctMixed}{}

\end{document}